# Mechanically-tunable bandgap closing in 2D graphene phononic crystals

*Jan N. Kirchhof[1,*] and Kirill I. Bolotin[1,*]*

[1] Department of Physics, Freie Universität Berlin, Arnimallee 14, 14195 Berlin, Germany

*jan.kirchhof@fu-berlin.de   *kirill.bolotin@fu-berlin.de

**Abstract**

**We present a tunable phononic crystal which can be switched from a mechanically insulating to a mechanically conductive (transmissive) state. Specifically, in our simulations for a phononic lattice under biaxial tension ($\sigma_{xx} = \sigma_{yy}$ = 0.01 N m$^{-1}$), we find a bandgap for out-of-plane phonons in the range of 48.8 – 56.4 MHz, which we can close by increasing the degree of tension uniaxiality ($\sigma_{xx}/\sigma_{yy}$) to 1.7. To manipulate the tension distribution, we design a realistic device of finite size, where $\sigma_{xx}/\sigma_{yy}$ is tuned by applying a gate voltage to a phononic crystal made from suspended graphene. We show that the bandgap closing can be probed via acoustic transmission measurements and that the phononic bandgap persists even after the inclusion of surface contaminants and random tension variations present in realistic devices. The proposed system acts as a transistor for MHz-phonons with an on/off ratio of 10$^5$ (100 dB suppression) and is thus a valuable extension for phonon logic applications. In addition, the transition from conductive to isolating can be seen as a mechanical analogue to a metal-insulator transition and allows tunable coupling between mechanical entities (e.g. mechanical qubits).**

# INTRODUCTION

Phononic crystals (PnCs) are artificial structures in which the periodic variation of material properties (e.g. stiffness, mass, or tension) give rise to a phononic band structure – in analogy to Bloch waves in crystalline solids on the atomic scale. In contrast to conventional solids, the parameters of the band structure can be broadly controlled via artificial patterning. Because of that, PnCs allow realizing analogues of fundamental solid state physics phenomena over a very large range of sizes (10 nm – 100 m) and frequencies (Hz - THz).[1,2] This ranges from quantum entanglement[3,4] to topological states[5,6] and negative refraction.[7] The ability to engineer phononic spectra gave rise to applications such as phononic shielding in ultracoherent mechanical resonators,[8–11] wave guiding[12,13] or thermal management.[14] Due to the much lower propagation speed of phonons compared to photons or electrons, PnCs are also promising candidates for quantum information technology based on guiding and storing mechanical motion, especially on length scales too small for photonic approaches.[6,15–18] Most of these applications and phenomena rely on phononic bandgaps, the range of frequencies where no phonons are allowed and mechanical motion is heavily damped.

The velocities of all phonons in a material depend on its tension $\sigma$. In conventional rigid PnCs, e.g. those fabricated using silicon nitride membranes ($SiN_x$), the built-in tension is determined during the growth step and cannot be tuned. As a result, it becomes challenging to couple a PnC to an external system, for example for processing and storing of quantum information.[19–21] In contrast, it has been recently demonstrated that the tension in much more flexible two-dimensional (2D) materials can be dynamically controlled by applying electrostatic pressure via an external gate electrode.[22–25] The resulting tunable (biaxial) tension allows broad tunability of the bandgap centre frequency.[23] Nevertheless, the hierarchy of the bands in the systems explored so far has not been affected by tension – i.e. a gapped system remained gapped at any tension level. The precise control of the bandgap size and thus the coupling strength between mechanical entities remains elusive.

Here, we show that the application of uniaxial tension to a PnC (in contrast to biaxial tension studied previously) changes the band hierarchy. Specifically, for a 2D phononic lattice patterned into a suspended graphene membrane under biaxial tension ($\sigma_{xx}/\sigma_{yy} = 1$), we observe a bandgap for out-of-plane phonons at any tension (e.g. 48.8 – 56.4 MHz at $\sigma = 0.01$ N m$^{-1}$), which disappears completely when the degree of tension uniaxiality ($\sigma_{xx}/\sigma_{yy}$) reaches 1.7. This can be seen as the observation of a mechanical analogue to a metal-insulator transition. Of course, the analogy is not complete. The chemical potential for the phononic system is zero rather than falling into the gap, as is the case for electrical insulators, which are described by the Fermi-Dirac statistics. Also, the analogy is only applicable to out-of-plane modes. These modes are especially relevant in phononic crystals made from 2D materials as they are easy to excite and detect. Nevertheless, the transition from a gapped to non-gapped phononic crystal shows many similarities to an actual metal-insulator transition in terms of transfer of energy and localisation of modes (see Supplementary Note 8). To control $\sigma_{xx}/\sigma_{yy}$, we propose a simple experimental geometry based on electrostatic gating and show that bandgap closing can be reached in experimentally feasible devices, which we probe via acoustic transmission studies. Our simulations show that applying a small gate voltage of ~8 V to the suspended graphene PnC is sufficient to close the phononic bandgap. Within in the bandgap region, the system functions as a mechanical transistor with an on/off ratio of $10^5$ (suppression of 100 dB) and can be used in phonon logic circuits. Furthermore, the ability to dynamically control the bandgap size allows to realize tunable coupling strength between mechanical entities e.g. two mechanical resonators acting as qubits. Finally, we investigate the challenges associated with the fabrication of 2D materials. We find that the mass of contaminants on top of the device must be smaller than ~4 times the weight of the suspended graphene and that the relative tension variation in the graphene must be smaller than 40% to observe a clear bandgap and its closing.

## **RESULTS**

**PnC design**

For the design of our tunable 2D phononic system we choose a honeycomb lattice (lattice constant *a*) of holes (diameter *d*), which provides a relatively broad and robust phononic bandgap for out-of-plane modes, while leaving a large fraction of the material untouched. The latter is crucial for making a PnC from fragile 2D materials. The honeycomb lattice also features an indirect phononic bandgap, which allows selective tuning of phononic bands via uniaxial tension, as we will see later. We select graphene as a suitable material for our PnC as it is the most conductive[26,27] and the strongest member of the family of 2D materials.[28] Our results are also applicable to other conductive 2D materials. The phononic pattern shows the same symmetry as the atomic lattice structure of graphene, with the difference that in our approach the unit cell is much larger and contains ~4·10[7] carbon atoms. We consider a free-standing PnC to allow mechanical tuning via out-of-plane pressure. Fabrication of such devices has recently been demonstrated by He-Ion beam milling.[23,29,30] To obtain the phononic band structure, we start by performing finite element method (FEM) simulations of the tension distribution within the conventional unit cell of the honeycomb lattice (Fig. 1a, top). We find tension hotspots in the thin ribbons and relaxed regions in the centre of the unit cell. This redistribution of tension occurs when holes are cut into the initially uniform membrane. In a next step, we use the first Brillouin zone (Fig. 1a, bottom) to calculate the phononic band structure along the high symmetry lines for an infinite lattice, as shown in Fig. 1b for $a = 1$ μm, $d/a = 0.5$ and a reasonable initial biaxial tension of $\sigma_{xx} = \sigma_{yy} = 0.01$ N m$^{-1}$.[23,31,32] For out-of-plane modes (solid lines) we find a bandgap between 48.8 and 56.4 MHz (blue shaded), in agreement with previous work.[23,30] These modes are qualitatively comparable to atomic scale flexural (ZA) phonon modes in graphene, but at much lower frequencies and for much smaller wave vectors. The entire phononic lattice behaves like a thin membrane with vibrational mode frequencies $f$ determined by the built-in tension ($f \sim \sqrt{\sigma}$), that also results in a linear dispersion for the flexural modes, instead of the quadratic behaviour expected for an unstrained 2D material.[33,34] Also, in agreement with previous work, we find that an uniform increase in tension ($\sigma_{xx}/\sigma_{yy} = 1$) leads to monotonic upscaling of both the top of valence ($f_{VB}$) and bottom of conduction band ($f_{CB}$) frequencies as shown in Fig. 1e (red). Here, the centre frequency of the bandgap follows a square root behaviour vs. tension, and the relative bandgap size ($\frac{f_{CB} - f_{VB}}{(f_{CB} + f_{VB})/2}$) remains constant.

**Bandgap closing for highly uniaxial tension**

Our next goal is to show that we can use uniaxial tension (unlike biaxial tension) to control the relative bandgap size and even completely close it. The phononic bandgap of a hexagonal lattice is indirect with the conduction band minima $f_{CB}$, located at the Γ point in momentum space and the valence band maxima $f_{VB}$, at a point along the ΓX line (Fig. 1b). Critically, uniaxial tension, in contrast to biaxial tension, produces different frequency scaling of the band structure at different points of the Brillouin zone. With increasingly uniaxial tension, $f_{VB}$ strongly upshifts in frequency while $f_{CB}$ is barely tension-dependent. As a result, the indirect bandgap of the phononic lattice acquires a strong tension-dependence. To quantify these changes, we determine the average tension components (after redistribution upon phononic pattering) $\sigma_{ij} = <\sigma_{ij}>$ and use the ratio $\sigma_{xx}/\sigma_{yy}$ as a metric for tension uniaxiality. For the honeycomb lattice with its initial tension distribution (as introduced above), $\sigma_{xx}/\sigma_{yy} = 1$. For an increased $\sigma_{xx}/\sigma_{yy} = 1.35$, we find increased tension in the areas stretched in the $x$ direction (Fig. 1c, inset). This is accompanied by a much more pronounced upshift of $f_{VB}$ compared to $f_{CB}$ and thus a reduced bandgap size (Fig. 1c). To give an intuitive understanding of this scaling behaviour, we look at the spatial shape of modes corresponding to $f_{VB}$ and $f_{CB}$. The mode $f_{CB}$ at the Γ point (Fig. 1b, left inset) resembles a standing wave along the $y$ direction, and it therefore does not depend strongly on tension in the $x$ direction. The mode corresponding to $f_{VB}$, between Γ and X (Fig. 1b, right inset), resembles a linear combination of standing waves in the $x$ and $y$ directions. The frequency of this mode however does depend on $\sigma_{xx}$. For a higher uniaxiality of 1.7 as shown in Fig. 1d, the tension distribution becomes even more distorted (Fig. 1d inset) and the lower branches ($f_{VB}$) overtake the upper ones ($f_{CB}$). At this point, the bandgap closes ($\frac{f_{CB}-f_{VB}}{(f_{CB}+f_{VB})/2} = 0$). In Supplementary Note 1, we provide extended band structure calculations showing the full extent of the Brillouin zone under uniaxial tension.

To summarize the results of bandgap tuning, in Fig. 1e we compare $f_{VB}$ and $f_{CB}$ vs. the total tension $\sigma_{total}$ for uniaxial (blue) and uniform biaxial (red) tension. For uniaxial tension, we see a closing of the bandgap at $\sigma_{total}/\sigma_0 = 1.6$ (corresponds to $\sigma_{xx}/\sigma_{yy} = 1.7$). In contrast, for biaxial tension scaling, the bandgap increases in absolute size with increased tension, while the relative bandgap size remains constant. Overall, by varying the tension uniaxiality, we find different scaling behaviour for different phononic bands along different directions, which allows us to dynamically tune the size of the bandgap.

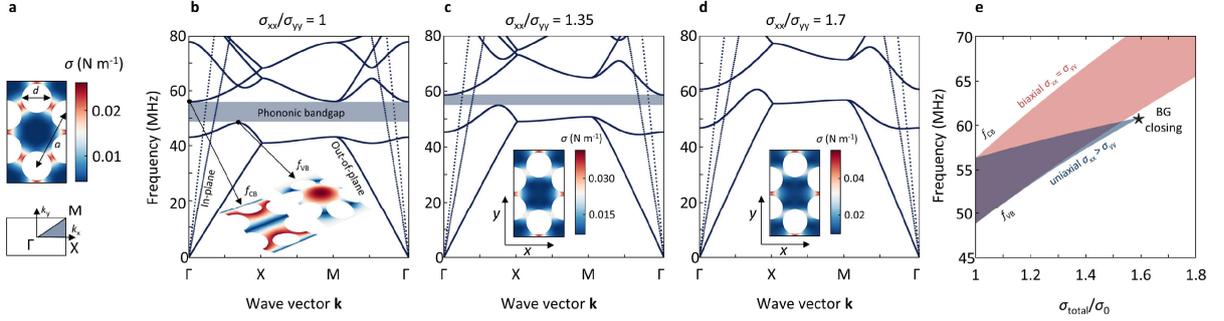

**Figure 1 | Phononic bandgap closing induced by uniaxial tension. a**, Unit cell of the honeycomb lattice with redistributed tension (top) and the corresponding first Brillouin zone (bottom). **b**, Phononic band structure for the unit cell shown in (a) under uniform tension ($\sigma_{xx} = \sigma_{yy} = 0.01$ N /mm$^{-1}$). For out-of-plane modes (solid lines) a clear phononic bandgap is visible (blue shaded region). The insets show the mode shape (displacement) within the unit cell at the points above and below the bandgap. **c,d**, Phononic band structure and tension distribution in the unit cell (insets) for $\sigma_{xx}/\sigma_{yy}$ of 1.35 and 1.7. With increasing uniaxiality in tension ($\sigma_{xx}/\sigma_{yy} > 1$), the phononic bands show different frequency scaling behavior along different high symmetry lines. At $\sigma_{xx}/\sigma_{yy} = 1.7$, the phononic bandgap closes. **e**, Phononic bandgap for biaxial (red) and uniaxial (blue) tension vs. total normalized tension. When the tension is increased biaxially ($\sigma_{xx} = \sigma_{yy}$), the bandgap centre frequency rises, and the bandgap width increases. On the contrary, uniaxial upscaling ($\sigma_{xx} > \sigma_{yy}$) leads to a bandgap closing.

Our next goal is to develop a design for an experimentally feasible device realizing the phononic bandgap closing. To accomplish this, three challenges need to be overcome. First, how can we probe the phononic bandgap in a realistic finite-size device? This is critical as the band structure calculations considered so far always assume an infinite phononic lattice. Second, how can we generate the highly uniaxial tension distribution needed to close the bandgap? Third, is it feasible to fabricate a sufficiently uniform PnC from experimentally available 2D materials? We now individually address each of these questions in the next sections.

**Bandgap probing in a finite-size device**

We probe our finite-size phononic system via acoustic transmissions measurements. In general, the transmission across a phononic system is determined by the density of available states at the relevant frequency which serves as a proxy for the phononic band structure. We design a transistor-style PnC with realistic dimensions of 9 μm x 28 μm (7 x 17 unit cells, unlike the infinite system considered in simplified simulations so far), in which instead of electrons we will determine the transmission of mechanical motion (Fig. 2a). At point A (excitation/source) mechanical motion is excited, which then can propagate through the PnC until it reaches point B (detection/drain). Drive and detection in such a device design can be experimentally realized by using either surface acoustic waves (SAW),[15] local gate electrodes[35] or two spatially separated laser beams[12] (blue, red Fig. 2a). Here, we concentrate without loss of generality on the last case. We define the transmission from area A to B as:

$$\text{Transmission}_{A \Rightarrow B}(f) = \frac{1}{T} \int_0^T \frac{\iint_A z(x,y,f,t)dA}{\iint_B z(x,y,f,t)dA} dt, \tag{1}$$

where $z(x, y, f, t)$ is the out of plane displacement of the suspended graphene with a period $T$ ($f = \frac{1}{T}$). The integration is over the illumination areas in points A and B. We concentrate on out-of-plane modes as they are controlled by the in-plane phononic pattern, show strong capacitive coupling to perpendicular electric fields from a gate electrode and are sensitive to interferometric readout. In Fig. 2b, we plot the transmission vs. frequency for the device shown in Fig. 2a. In the region below the fundamental resonance, the stop band (< 5 MHz), we find strongly supressed transmission. Towards higher frequencies, we find multiple closely spaced sharp peaks, which correspond to higher order resonances of the device. As the frequency increases further, the transmission is more and more dominated by the phononic band structure, and we observe broad "bands" rather than individual resonance modes. The transmission suddenly drops by an average of 5 orders magnitude in the expected bandgap region between 48.5 and 56.5 MHz (blue shaded). The non-zero transmission inside the bandgap is related to finite-size effects captured by our model. Above the bandgap the transmission recovers and remains close to 1. The frequency range of the bandgap extracted from transmission simulations matches well with the bandgap from band structure calculations (comp. Fig. 1b). To summarize, we can use acoustic transmission studies to probe the phononic bandgap in finite-size devices. Furthermore, transmission of mechanical motion across the device in the bandgap region is strongly suppressed and, in analogy to an electronic system, the system can be considered a mechanically insulating.

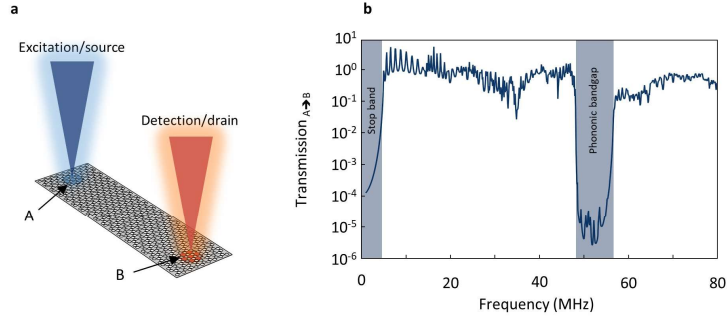

**Figure 2 | Probing the band structure via transmission studies in a finite-size phononic crystal. a**, Transmission geometry for a rectangular phononic device. At point A mechanical motion is excited by a frequency modulated laser (blue). The vibrational wave travels through the device and is detected at point B by a second laser spot (red). **b**, Transmission from A to B vs. excitation frequency for the device shown in (a). A clear bandgap region is visible (blue shaded) where transmission of mechanical motion through the device is suppressed by ~$10^5$.

**Uniaxial tension engineering**

After finding the phononic bandgap closing in band structure calculations at a tension uniaxiality of $\sigma_{xx}/\sigma_{yy} = 1.7$ and establishing transmission studies as a suitable approach to probe the bandgap, we now aim to produce the required tension distribution – and hence the bandgap closing – in a realistic device of finite size. Our key idea is to apply electrostatic pressure to a suspended rectangular device (Fig. 3a) with non-unity aspect ratio ($W/L$). In this case the induced tension is larger along the direction of the smaller spatial dimension ($x$ in Fig. 3a).[36] We model the membrane as clamped at its perimeter. Electrostatic pressure $p_{el}$ is generated by applying a gate voltage ($V_{gate}$) between the highly conductive graphene and a gate electrode separated from it by distance $d$:

$$p_{el} = \frac{\epsilon_0}{2}\left(\frac{V_{gate}}{d}\right)^2, \qquad (2)$$

where $\epsilon_0$ is the vacuum permittivity. We assume $d = 300$ nm, a typical oxide thickness for Si/SiO$_2$ substrates used for 2D materials. At zero gate voltage, corresponding to zero pressure, the membrane is uniformly tensed ($\sigma_{xx} \approx \sigma_{yy}$). The tension distribution inside the center of the phononic device is plotted in Fig. 3b. With applied pressure the degree of tension uniaxiality $\sigma_{xx}/\sigma_{yy}$ increases and the distribution of tension becomes rotationally asymmetric (Fig. 3c). For $p_{el} = 3$ kPa (8 V), $\sigma_{xx}/\sigma_{yy}$ reaches 1.7 and we thus expect the bandgap closing to occur. The generated tension distribution also matches the prediction for the bandgap closing from our band structure calculations – compare dashed outline in Fig. 3c with the inset of Fig. 1d. In Fig. 3d we summarize the results of tension engineering for our finite-size system in a phase diagram, where we plot $\sigma_{xx}/\sigma_{yy}$ vs. applied pressure vs. aspect ratio. When $\sigma_{xx}/\sigma_{yy}$ reaches the critical value of 1.7 (dashed line), we expect bandgap closing according to our band structure calculations for the infinite lattice. This line can therefore be viewed as a boundary separating a mechanically insulating from a mechanically conductive (transmissive) state. We see that the conductive state is reached at lowest applied pressure for an aspect ratio of $W/L = 0.32$.

Next, we calculate the transmission spectra for applied pressures of 0 and 5 kPa (Fig. 3e). While we find a clear bandgap (and higher order harmonics) for the un-pressured state (blue), the bandgap completely vanishes with applied pressure (red, 5 kPa), confirming the expected bandgap closing for a finite-size phononic crystal. The system is now transmissive and mechanically conductive. Continuing the analogy between phononic and electronic devices, our system can be viewed as a mechanical transistor for MHz phonons with an on/off ratio of ~$10^5$ (100 dB suppression). This corresponds to 6dB suppression per unit cell.

In Fig. 3f, we show combined results from multiple pressures by plotting $f_{VB}$ and $f_{CB}$ for the rectangular device ($W/L = 0.32$, blue) and a circular reference device (red). In accordance with previous simplified calculations (Fig. 1e), we see that the bandgap $\frac{f_{CB}-f_{VB}}{(f_{CB}+f_{VB})/2}$ gradually decreases in size with applied pressure for the rectangular device. The applied pressure increases $\sigma_{xx}/\sigma_{yy}$ and drives the system towards the bandgap closing. In contrast, the circular reference device for which we expect entirely biaxial tension tuning ($\sigma_{xx} \approx \sigma_{yy}$) exhibits a clear bandgap up to 30 kPa (see Supplementary Note 3). To better relate our results to the phononic band structure calculations, we take the average tension values ($\sigma_{xx}$, $\sigma_{yy}$ and $\sigma_{total}$) from the finite-size system under pressure as input for our infinite model and plot the expected bandgap regions in Fig. 3f (red and blue shaded). While we find comparable behaviour, the bandgap closing however occurs at somewhat higher pressures. This is likely due to boundary-related disorder that is excluded in the infinite model. We extract the average strain from our simulations and obtain $\varepsilon = 0.24\%$ for an applied pressure of 10 kPa. This is well below the onset of phonon instabilities[37] or graphene's breaking strain.[28] To summarize, we find bandgap closing for a highly uniaxial tension distribution generated by applying electrostatic pressure in a realistic finite-size device with optimized geometry. This allows us to change the state of a PnC from mechanically insulating to conductive by simply applying a gate voltage.

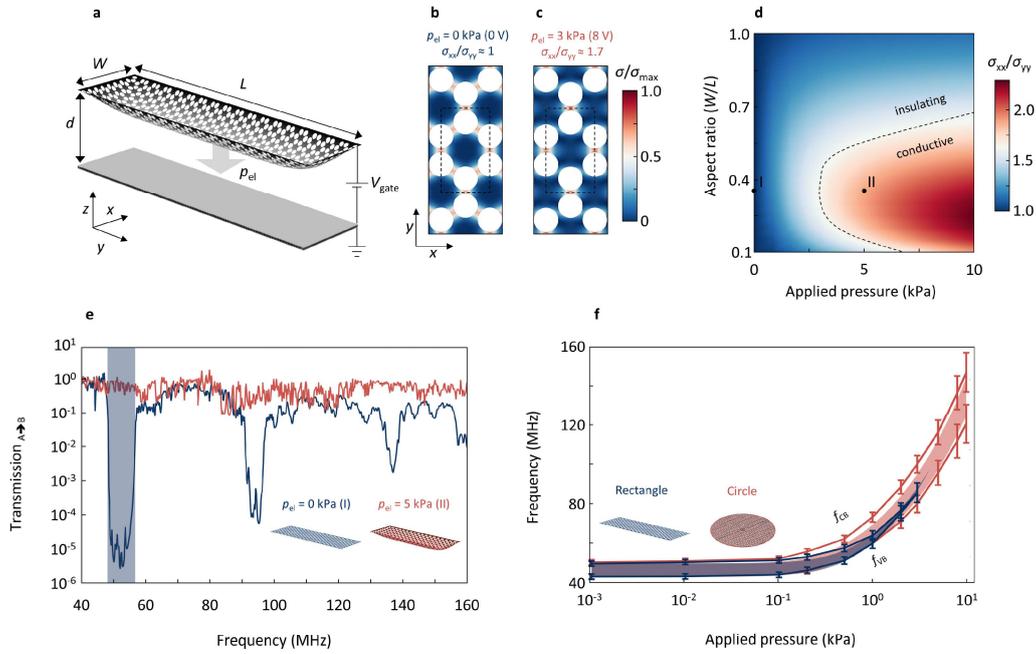

**Figure 3 | Uniaxial tension engineering in a finite-size phononic system. a,** Sketch of a finite-size system phononic device, which is mechanically deformed under electrostatic pressure, $p_{el}$, generated by a gate electrode below the graphene. **b,c,** Spatial tension distribution in the centre of the device with and without applied pressure. The dashed lines indicate the unit cell of the lattice. **d,** Mechanical phase diagram: Tension uniaxiality ($\sigma_{xx}/\sigma_{yy}$) vs. pressure vs. device aspect ratio ($W/L$). The dotted line corresponds to $\sigma_{xx}/\sigma_{yy} = 1.7$, the degree in uniaxiality needed to close the bandgap. **e,** Transmission for a device with an aspect ratio of 0.32 for $p_{el}$ = 0 kpa kPa (blue) and 5 kPa (red). The initially pronounced bandgap vanishes with applied pressure. **f,** Extracted valence band maximum ($f_{VB}$) and conduction band minimum ($f_{CB}$) vs. applied pressure for a rectangular device (blue, shown in (e)) and a circular device (red) as reference for uniform scaling ($\sigma_{xx}/\sigma_{yy} \approx 1$). For the rectangular device the bandgap closing occurs at 3 kPa, whereas the circular device maintains a bandgap over the entire range of applied pressures. The error bars depict the reading error of the simulation results, and the shaded areas correspond to the phononic bandgap extracted from band structure calculations.

**Fabrication related challenges**

Having demonstrated large frequency tunability as well as phononic bandgap closing in graphene PnCs, we now want to assess the fabrication challenges associated with 2D materials. We therefore investigate the effect on the phononic bandgap for the two most common forms of disorder in 2D materials: surface contamination and random tension variations.

Perhaps the most widespread sources of contamination are "islands" of residues on top of the graphene. To simulate these added pieces of mass, we choose Polydimethylsiloxan (PDMS) as a typical polymer often used for transfer of 2D materials, and randomly place the pieces on the graphene membrane (Fig. 4a). At a thickness of 18 nm and a diameter of 4 μm, each piece has the same weight as the entire clean resonator. Next, we focus on the bandgap region and plot transmission vs. frequency for various amounts of added mass (Fig. 4b). Even for three added pieces (red), we still observe weak signatures of the phononic bandgap and conclude that the combined mass density of graphene and contamination must be on the order of $\rho_{2D} \leq 4\rho_{\text{graphene}}$. Values below this threshold have been observed in some graphene resonators in literature.[32,38] We also test the effect of a uniform film of PDMS on the phononic device and still find a clear bandgap (see Supplementary Note 4).

The second potential threat for breaking the phononic order are random tension variations in the suspended membrane commonly observed in both patterned and unpatterned graphene membranes.[31] To model this effect, we generate disorder based on a superposition of randomized plane waves (details in Supplementary Note 5). We take into account variations down to ¼ of the lattice constant of the phononic pattern. Two generated spatial tension distributions for small and large disorder are shown in Fig. 4c. The disorder strength is parametrized by the standard deviation of the distribution, $\sqrt{\text{Var}(\sigma)}/\sigma_0$ (see insets). We now calculate the transmission through the phononic device as a function of disorder strength. As shown in Fig. 4d, we find a gradual smearing out of the bandgap with increasing disorder. Above an estimated critical value of $\sqrt{\text{Var}(\sigma)}/\sigma_0 \approx 0.4$, the bandgap is no longer clearly distinguishable. If we compare this threshold to experimental values derived from Raman spectroscopy,[39–41] we find similar spreads in tension. We also investigate the effect of variations in hole size on the phononic bandgap and find it to be robust for the level of disorder seen in realistic devices (see Supplementary Note 6). We conclude that it is challenging but possible to fabricate sufficiently uniform suspended devices. If, however more uniform samples are needed, we propose using thin multilayers of graphene, for which we find a bandgap up to a thickness of ~200 layers (see

Supplementary Note 7). For multilayer devices, we need larger pressures to induce the bandgap closing, but commonly used SiO$_2$/Si (300 nm) substrates allow applying ~100 V gate voltage before dielectric breakdown occurs, which translates to ~50 kPa (sufficient to induce the bandgap closing on multilayer devices). Overall, fabricating a PnC from suspended graphene with a pronounced bandgap is feasible.

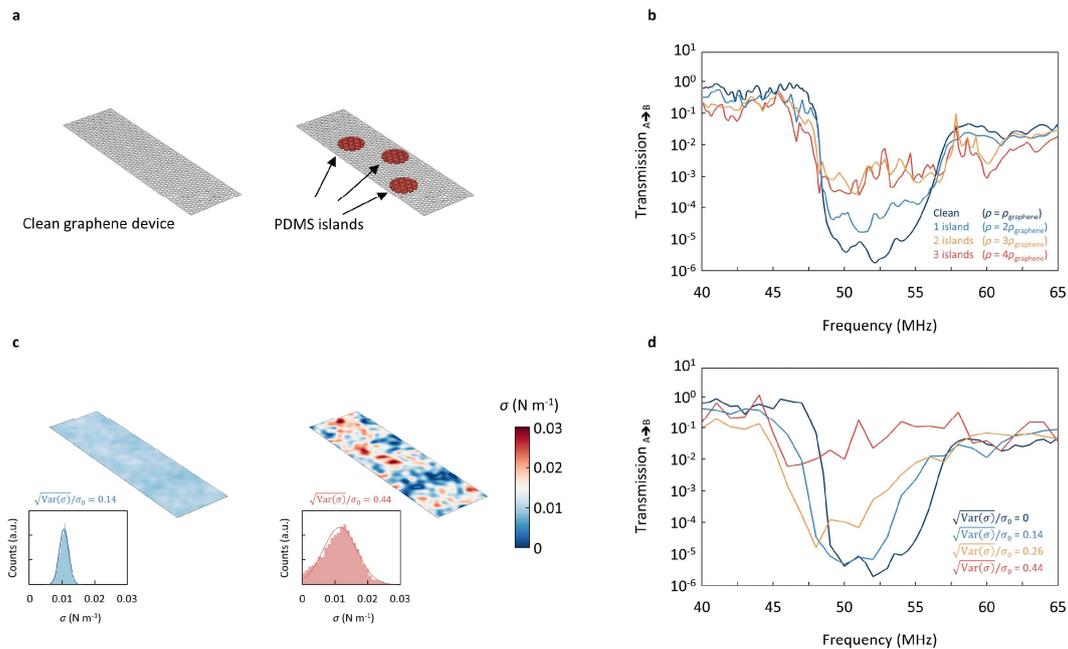

**Figure 4 | The effect of disorder on the phononic bandgap. a,** A phononic device with and without surface contamination. **b,** Phononic bandgap vs. added mass. With increasing degree of contamination, the bandgap smears out, yet remains visible up to areal mass density of $\rho_{2D} \approx 4\rho_{graphene}$. **c,** Graphene membrane before patterning with small and large tension disorder. The insets show the histograms used to extract the disorder strength: left, $\sqrt{\text{Var}(\sigma)}/\sigma_0 = 0.14$ and right, $\sqrt{\text{Var}(\sigma)}/\sigma_0 = 0.44$. **d,** Phononic bandgap vs. tension disorder. At a critical relative variation in tension $\sqrt{\text{Var}(\sigma)}/\sigma_0 \approx 0.40$ the bandgap vanishes.

## Discussion

We have demonstrated the manipulation of the phononic band structure by using uniaxial tension engineering and found closing of a phononic bandgap at $\sigma_{xx}/\sigma_{yy} = 1.7$. This transition from a mechanically insulating to a conductive state may be regarded as the mechanical analogue of a metal-insulator-transition. In a finite-size device, we can generate the required uniaxial tension distribution by applying a voltage of ~8 V to a gate electrode and observe vanishing of the phononic bandgap in transmission studies. This device can be considered a phononic counterpart to a field effect transistor, with acoustic transmission measurements at the bandgap frequency taking the role of electrical transport. Furthermore, we discuss the feasibility of fabricating such a device with commonly used methods and extract a critical value for surface contamination ($\rho_{2D} \leq 4\rho_{graphene}$) and tension variations ($\sqrt{\mathrm{Var}(\sigma)}/\sigma_0 \approx 0.40$).

The proposed system acts as a phononic transistor that can be used for phonon logic in the MHz range and invites realisation of a variety of logic gates as a next step. By varying the lattice constant $a$, the phononic system can be engineered to function in a broad range of frequencies from ~10 MHz to ~1 GHz. In addition, the proposed device design can serve as a switch controlling the coupling between two remote systems, e.g. mechanical resonators acting as qubits.[19–21] This in principle also allows tunable dispersive readout of qubits via mechanical resonators. The proposed bandgap closing also makes it possible to control the phononic shielding of ultracoherent defect modes from the environment and therefore allows to dynamically study dissipation mechanisms as shown in Supplementary Note 8. Finally, following the analogy between phononic and electronic crystals invites the consideration of analogues to other, more complex condensed matter physics phenomena, e.g. the quantum hall effect, Mott insulator transition, and topological phase transitions.

# Methods

**FEM Simulations**

For the finite element modelling, we use COMSOL Multiphysics (Version 5.5) and assume the following material parameters for monolayer graphene: Young's modulus $E_{2D}$ = 1.0 TPa[28], Poisson's ratio of $v$ = 0.15, thickness of $h$ = 0.335 nm and a density of $\rho = \frac{\rho_{2D}}{h}$ = 2260 kg m$^{-3}$. For details, see Supplementary Note 1-8.

**Acknowledgements**

This work was supported by Deutsche Forschungsgemeinschaft (DFG, German Research Foundation, project-ID 449506295 and 328545488), CRC/TRR 227 and ERC Starting grant no. 639739. We thank Yuefeng Yu, Aleksei Tsarapkin, Victor Deinhart and Katja Höflich for the fabrication of a graphene phononic crystal reference sample.

**Author contributions**

J.N.K and K.I.B. conceived the idea and wrote the manuscript. J.N.K. performed the FEM simulations.

**COMPETING INTERESTS**

The authors declare no competing interests.

**DATA AVAILABILITY**

The data that support the findings of this study are available from the corresponding author upon reasonable request.

**CODE AVAILABILITY**

Upon request, authors will make available any previously unreported computer code or algorithm used to generate results that are reported in the paper and central to its main claims.

# References

1. Kushwaha, M. S., Halevi, P., Dobrzynski, L. & Djafari-Rouhani, B. Acoustic band structure of periodic elastic composites. *Phys. Rev. Lett.* **71**, 2022–2025 (1993).

2. Maldovan, M. Sound and heat revolutions in phononics. *Nature* **503**, 209–217 (2013).

3. Thomas, R. A. *et al.* Entanglement between distant macroscopic mechanical and spin systems. *Nat. Phys.* 1–6 (2020). doi:10.1038/s41567-020-1031-5

4. Riedinger, R. *et al.* Remote quantum entanglement between two micromechanical oscillators. *Nat. 2018 5567702* **556**, 473–477 (2018).

5. Mousavi, S. H., Khanikaev, A. B. & Wang, Z. Topologically protected elastic waves in phononic metamaterials. *Nat. Commun.* **6**, 1–7 (2015).

6. Pirie, H., Sadhuka, S., Wang, J., Andrei, R. & Hoffman, J. E. Topological Phononic Logic. *Phys. Rev. Lett.* **128**, (2022).

7. He, H. *et al.* Topological negative refraction of surface acoustic waves in a Weyl phononic crystal. *Nat. 2018 5607716* **560**, 61–64 (2018).

8. Tsaturyan, Y., Barg, A., Polzik, E. S. & Schliesser, A. Ultracoherent nanomechanical resonators via soft clamping and dissipation dilution. *Nat. Nanotechnol.* **12**, 776–783 (2017).

9. Ghadimi, A. H. *et al.* Elastic strain engineering for ultralow mechanical dissipation. *Science (80-. ).* **360**, 764–768 (2018).

10. Yu, P. L. *et al.* A phononic bandgap shield for high-Q membrane microresonators. *Appl. Phys. Lett.* **104**, 23510 (2014).

11. Li, F., Liu, J. & Wu, Y. The investigation of point defect modes of phononic crystal for high Q resonance. *J. Appl. Phys.* **109**, 124907 (2011).

12. Wang, Y., Lee, J., Zheng, X. Q., Xie, Y. & Feng, P. X. L. Hexagonal Boron Nitride Phononic Crystal Waveguides. *ACS Photonics* **6**, 3225–3232 (2019).

13. Otsuka, P. H. *et al.* Broadband evolution of phononic-crystal-waveguide eigenstates in

real- and k-spaces. *Sci. Rep.* **3**, 1–5 (2013).

14. Yang, L., Chen, J., Yang, N. & Li, B. Significant reduction of graphene thermal conductivity by phononic crystal structure. *Int. J. Heat Mass Transf.* **91**, 428–432 (2015).

15. Gustafsson, M. V. *et al.* Propagating phonons coupled to an artificial atom. *Science (80-. ).* **346**, 207–211 (2014).

16. Kumar, S. *et al.* Temperature-Dependent Nonlinear Damping in Palladium Nanomechanical Resonators. *Nano Lett.* acs.nanolett.1c00109 (2021). doi:10.1021/acs.nanolett.1c00109

17. Shin, H. *et al.* Control of coherent information via on-chip photonic-phononic emitter-receivers. *Nat. Commun.* **6**, 1–8 (2015).

18. Zivari, A. *et al.* On-chip distribution of quantum information using traveling phonons. (2022).

19. Navarathna, A. & Bowen, W. P. Good vibrations for quantum computing. *Nat. Phys. 2022* 1–2 (2022). doi:10.1038/s41567-022-01613-z

20. Luo, G. *et al.* Strong indirect coupling between graphene-based mechanical resonators via a phonon cavity. *Nat. Commun.* **9**, (2018).

21. Cirac, J. I., Zoller, P., Kimble, H. J. & Mabuchi, H. Quantum State Transfer and Entanglement Distribution among Distant Nodes in a Quantum Network. *Phys. Rev. Lett.* **78**, 3221–3224 (1997).

22. Hatanaka, D., Bachtold, A. & Yamaguchi, H. Electrostatically Induced Phononic Crystal. *Phys. Rev. Appl.* **11**, 1 (2019).

23. Kirchhof, J. N. *et al.* Tunable Graphene Phononic Crystal. *Nano Lett.* **21**, 2174–2182 (2021).

24. Zhang, Z.-D., Cheng, C., Yu, S.-Y., Lu, M.-H. & Chen, Y.-F. Electrically Tunable Elastic Topological Insulators Using Atomically Thin Two-Dimensional Materials Pinned on Patterned Substrates. *Phys. Rev. Appl.* **15**, 034015 (2021).

25. Zhang, Q. H. *et al.* Graphene-Based Nanoelectromechanical Periodic Array with Tunable Frequency. *Nano Lett.* **21**, 8571–8578 (2021).

26. Novoselov, K. S. *et al.* Two-dimensional gas of massless Dirac fermions in graphene. *Nature* **438**, 197–200 (2005).

27. Castro Neto, A. H., Guinea, F., Peres, N. M. R., Novoselov, K. S. & Geim, A. K. The electronic properties of graphene. *Rev. Mod. Phys.* **81**, 109–162 (2009).

28. Lee, C., Wei, X., Kysar, J. W. & Hone, J. Measurement of the elastic properties and intrinsic strength of monolayer graphene. *Science (80-. )*. **321**, 385–388 (2008).

29. Deinhart, V. *et al.* The patterning toolbox FIB-o-mat: Exploiting the full potential of focused helium ions for nanofabrication. *Beilstein J. Nanotechnol.* **12**, 304–318 (2021).

30. Kovalchuk, S., Kirchhof, J. N., Bolotin, K. I. & Harats, M. G. Non-Uniform Strain Engineering of 2D Materials. *Israel Journal of Chemistry* (2022). doi:10.1002/ijch.202100115

31. Nicholl, R. J. T. *et al.* The effect of intrinsic crumpling on the mechanics of free-standing graphene. *Nat. Commun.* **6**, 8789 (2015).

32. Chen, C. *et al.* Performance of monolayer graphene nanomechanical resonators with electrical readout. *Nat. Nanotechnol.* **4**, 861–867 (2009).

33. Bonini, N., Garg, J. & Marzari, N. Acoustic phonon lifetimes and thermal transport in free-standing and strained graphene. *Nano Lett.* **12**, 2673–2678 (2012).

34. Liu, F., Ming, P. & Li, J. Ab initio calculation of ideal strength and phonon instability of graphene under tension. *Phys. Rev. B - Condens. Matter Mater. Phys.* **76**, 064120 (2007).

35. Cha, J. & Daraio, C. Electrical tuning of elastic wave propagation in nanomechanical lattices at MHz frequencies. *Nat. Nanotechnol.* **13**, 1016–1020 (2018).

36. Kovalchuk, S. *et al.* Neutral and charged excitons interplay in non-uniformly strain-engineered WS2. *2D Mater.* **7**, 35024 (2020).

37. Liu, F., Ming, P. & Li, J. Ab initio calculation of ideal strength and phonon instability of graphene under tension. *Phys. Rev. B - Condens. Matter Mater. Phys.* **76**, 1–7 (2007).

38. Bunch, J. S. *et al.* Impermeable atomic membranes from graphene sheets. *Nano Lett.* **8**, 2458–2462 (2008).

39. Couto, N. J. G. *et al.* Random strain fluctuations as dominant disorder source for high-

quality on-substrate graphene devices. *Phys. Rev. X* **4**, 1–13 (2014).

40. Neumann, C. *et al.* Raman spectroscopy as probe of nanometre-scale strain variations in graphene. *Nat. Commun.* **6**, 8429 (2015).

41. Colangelo, F. *et al.* Mapping the mechanical properties of a graphene drum at the nanoscale. *2D Mater.* **6**, (2019).

# Supplementary Information: Mechanically-tunable bandgap closing in 2D graphene phononic crystals

*Jan N. Kirchhof[1*] and Kirill I. Bolotin[1*]*

[1] Department of Physics, Freie Universität Berlin, Arnimallee 14, 14195 Berlin, Germany

*jan.kirchhof@fu-berlin.de   *kirill.bolotin@fu-berlin.de

**Supplementary Note 1: Band structure calculations for an infinite lattice**

The infinite model for the phononic band structure calculations is based on two studies within the same FEM simulation model (Comsol Multiphysics 5.5). In the first step we simulate the tension redistribution upon pattering (stationary study). We then use the calculated tension distribution as input for a second study step (eigenfrequency study). In this step, we parameterize the x and y component of the k vector to cover the 1.BZ, and implement them into the model as periodic boundary conditions (Floquet) along the outline of the unit cell. We then calculate and plot the eigenfrequencies, what gives us the band structure. For more details see Ref. [1]. To assure that our simulations properly capture the phononic band gaps, we extend our calculations to more high symmetry points. For the unit cell shown in Supplementary Figure 1a, we plot the extended band structure for $\sigma_{xx}/\sigma_{yy} = 1$ and $\sigma_{xx}/\sigma_{yy} = 1.7$ inside the 1.Brillouin zone, see Supplementary Figure 1b,c. We find that the valence band maximum between Γ and X and the conduction band minimum at X, and it is therefore sufficient to focus on the high symmetry points shown in the main paper.

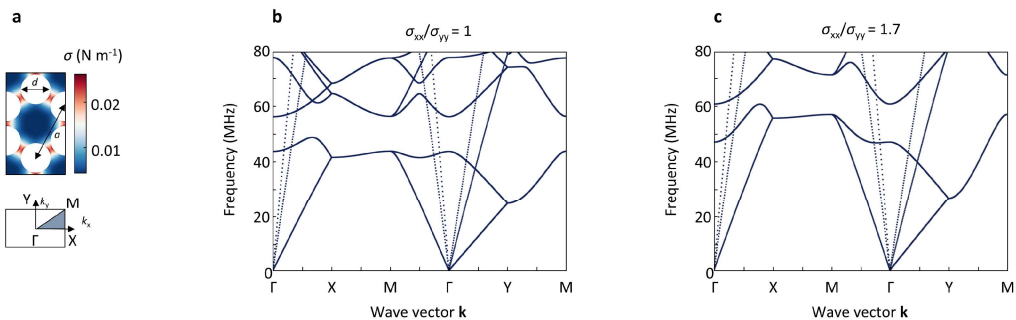

**Supplementary Figure 1 | Extended band structure. a**, Unit cell of the honeycomb lattice with redistributed tension (top) and the corresponding first Brillouin zone (bottom). **b,c**, Extended phononic band structure for the unit cell shown in (a) with entirely uniform tension ($\sigma_{xx}/\sigma_{yy}$ =1.) and implemented uniaxial tension $\sigma_{xx}/\sigma_{yy}$ =1.7.

**Supplementary Note 2: Transmission studies**

To perform our transmission studies, we use a pre-stressed frequency domain study. In this study the phononic device is clamped along its perimeter (Supplementary Figure 2) and in a first study step we again calculate the tension redistribution upon patterning. In a second step, we add a time depended pressure in z direction at area A, which simulates an optothermal drive. For the transmission study, we then sweep the frequency of this time depended perturbation and calculate the response of the entire geometry (compare Eq. 1 main text). For this study step, we add isotropic damping ($\eta = 0.01$) to the graphene, which reproduces the quality factors ($Q \sim 100$) typically observed for graphene resonators at room temperatures. To simulate the effect of electrostatic pressure to the transmission studies, we add a boundary load to the entire device (including A and B) in z-direction and repeat the frequency sweep.

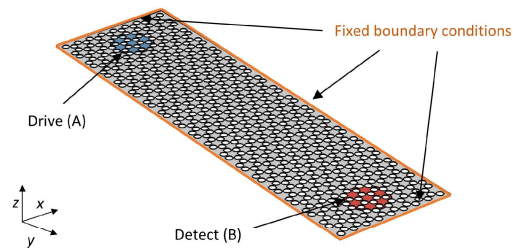

**Supplementary Figure 2 | Transmission geometry.** The phononic device is clamped at its perimeter via fixed boundary conditions and motion is excited at A (blue) and detected at B (red).

**Supplementary Note 3: Transmission circular reference device**

As a comparison to the presented bandgap closing in the main text we perform transmission studies on a circular device under applied pressure which is excited at its centre and probed on the outside. The device geometry and resulting transmission spectra are shown in Supplementary Figure 3a,b, where we find clear bandgap features (shaded area) with (red) and without (blue) applied pressure. In circular devices we find band gap features up to at least 30 kPa applied pressure in agreement with previous work.[1]

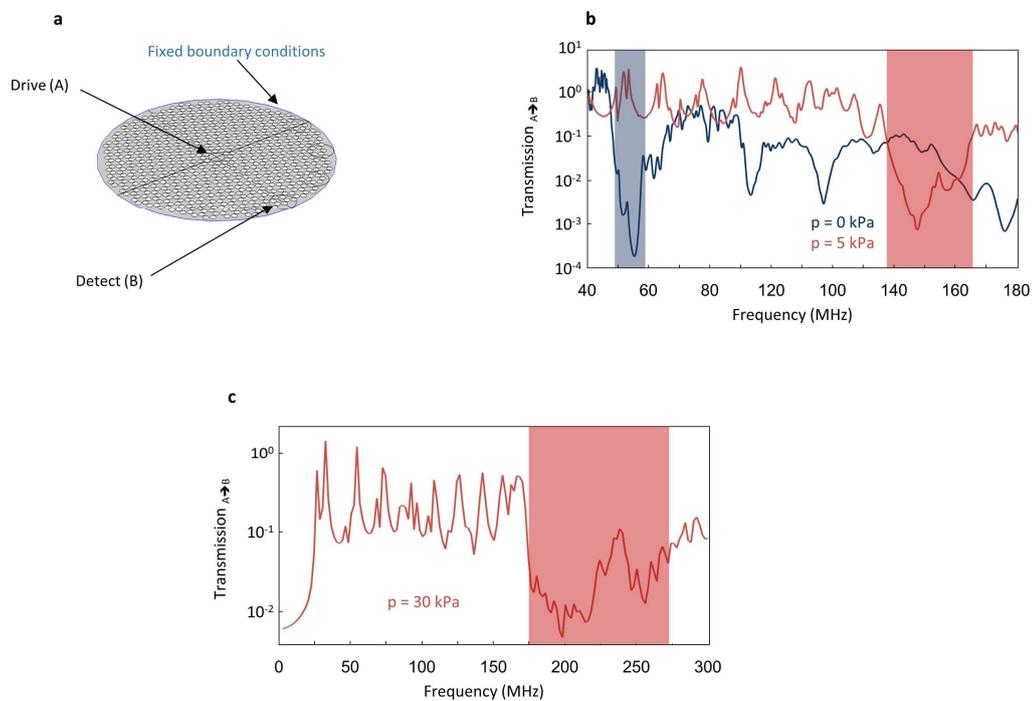

**Supplementary Figure 3 | Transmission circular reference device. a**, Transmission geometry for a rectangular phononic device. At point A mechanical motion is excited by a frequency modulated laser, which than travels through the device and is detected at point B by a second laser spot. **b,c** Transmission from A to B vs. excitation frequency for the device shown in (a) for 0, 5 and 30 kPa applied pressure. A clear bandgap region is visible for all cases.

**Supplementary Note 4: Phononic bandgap vs. uniform residues**

In addition the added pieces of mass (Figure 4 main text), we also investigate the effect of a uniform layer of resist, which could be present on a device after thermal annealing. In Supplementary Figure 4 we plot transmission for a device made from clean graphene (blue) and one contaminated with a 3 nm layer of PDMS (green). For both cases we find a clear bandgap, the added PDMS however causes a downshift in frequency as the entire device becomes heavier. Also the bandgap is slightly less pronounced, but still clearly noticeable. For PDMS we assume a Young's modulus of 0.75 MPa, a Poisson's ratio of 0.49 and a density of 970 kg m$^{-3}$.

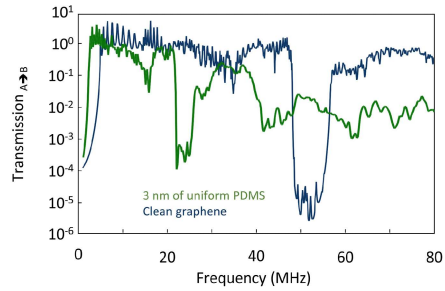

**Supplementary Figure 4 | Phononic bandgap vs. uniform residues.** Transmission vs. frequency for a phononic device made from clean graphene (blue) and with a 3 nm uniform layer of PDMS residues (green).

**Supplementary Note 5: Phononic bandgap vs. tension disorder**

To represent a random but smooth enough spatial tension distribution in our devices we use a sum of plane waves with randomized amplitude $a(m,n)$ (between -1 and 1) and phase $\phi(m,n)$ (between 0 and π) of each mode:

$$\sigma(x,y) = p \sum_{m=-M}^{M} \sum_{n=-N}^{N} a(m,n) \cos(2\pi(mx + ny) + \phi(m,n)) \quad (1)$$

The factor $p$ controls the disorder strength.

**Supplementary Note 6: Phononic bandgap vs. hole size variations**

We also investigate the effect of disorder within the phononic pattern in terms of a) variations of the radius of individual holes (so that the holes are not perfectly round) or b) size variations between the different perfectly circular holes that make up the phononic crystal.

Before we simulate the effect of a) and b) on the phononic bandgap, we need to get a feeling for the experimentally relevant disorder in patterned phononic crystals. To this end, we look at a real phononic crystal made from ~5 layer thick graphene (Supplementary Figure 5). We patterned this sample using the honeycomb lattice described in the manuscript by focused ion beam milling (FIB). The spatial resolution of this process is in the order of ~20 nm, which is compared to the hole diameter (500 nm) rather small, and we expect high precision and reproducible hole diameters. And indeed, we find a regular patterned phononic crystal as shown in Supplementary Figure 5.

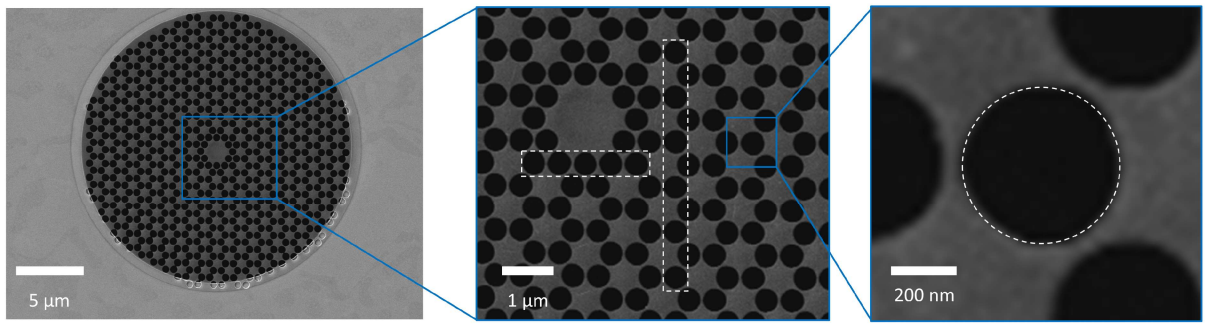

**Supplementary Figure 5 | Circular prototype device made from 5 Layer graphene for experimental work.** The phononic pattern is regular and the variations in hole size are small.

Next, we study the effect of non-circular holes (a) in our infinite model. We start by replacing two holes in the unit cell by ellipses (90° rotated to each other, eccentricity: $e = \sqrt{1 - \left(\frac{d_1}{d_2}\right)^2} = 0.574$, as shown in Supplementary Figure 6c) and calculate the corresponding band structure (Supplementary Figure 6d). If we compare this to the reference band structure obtained with perfect holes (Supplementary Figure 6a,b), we find that some bands split into closely lying sub-bands. Nevertheless, the bandgap remains. No change in size or position of the bandgap is noticeable. To go a step further, we try to capture variations in hole radius of individual holes comparable to the sample shown above (Supplementary Figure 5 right). To do so we implement exaggerated "wobbly" holes into the unit cell (Supplementary Figure 6e). Again, we find no major impact on the bandgap region (Supplementary Figure 6e) and thus the robustness of our results.

To check the effect of variations in hole size between different holes (b), we model a structure were the size of randomly picked holes ("defects") is decreased or increased by 10%. In Supplementary Figure 6g, we show the used structure with a defect density ($\frac{N_{\text{defect}}}{N_{\text{holes}}}$) of 12.5%. If we zoom in on a device with larger defect density of $\frac{N_{\text{defect}}}{N_{\text{holes}}} = 50\%$, the variations in hole size are clearly visible (see Supplementary Figure 6h). Next, we simulate the transmission for devices with defect densities from 0 to 50% (Supplementary Figure 6i). We find that the bandgap clearly persists to a defect density of at least 50%. This level of disorder is clearly much higher than that in experimentally achieved devices (Supplementary Figure 5). We therefore conclude that we can fabricate the devices ordered enough to observe the phenomena we study in the main text.

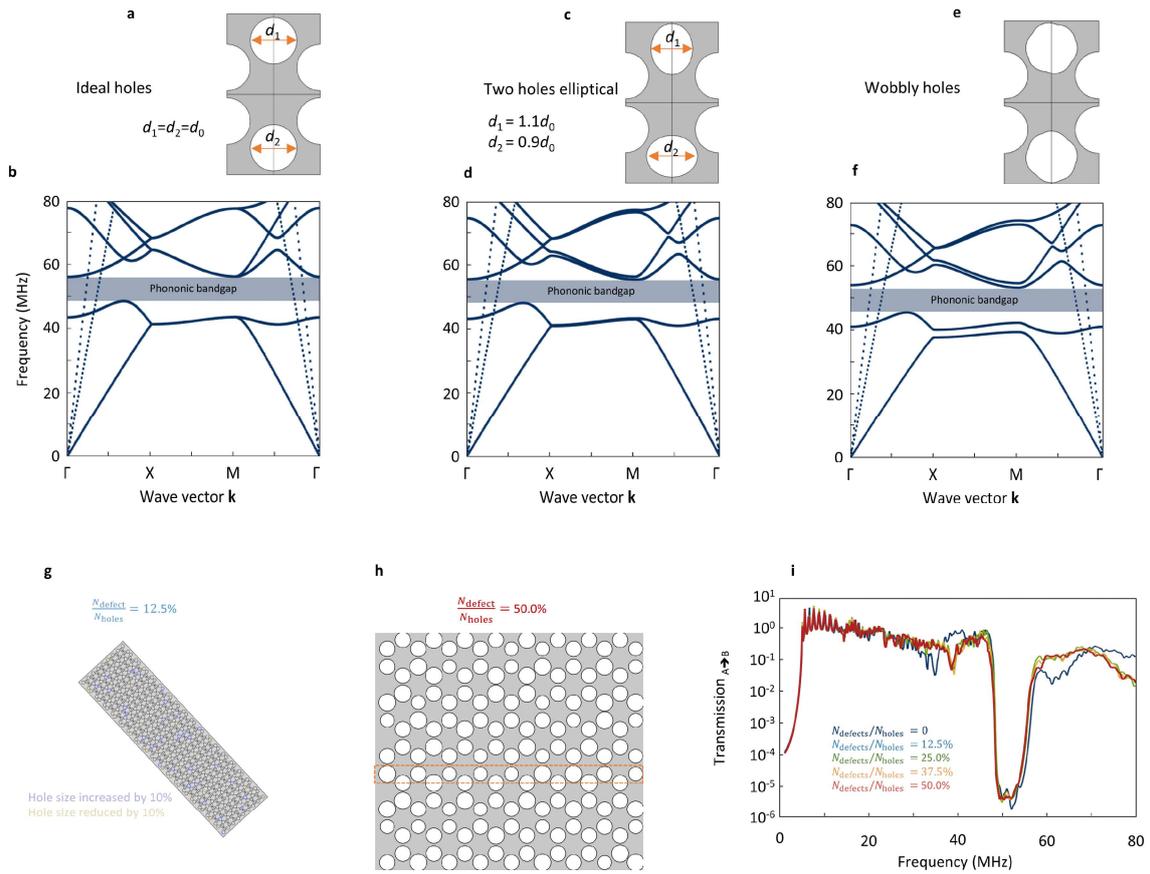

**Supplementary Figure 6 | Phononic bandgap vs. hole size variations a,b** Unit cell and corresponding band structure calculation for ideal holes (same as in main text Supplementary Figure 1b) **c,d** Unit cell with two elliptical holes and corresponding band structure calculation. The phononic bandgap remains almost unchanged. **e,f** Unit cell with two wobbled holes and corresponding band structure calculation The phononic bandgap remains almost unchanged. **g,h** Finite model with different degree of lattice defects (hole diameter variations) implemented. The blue (yellow) marking indicates holes, which will be increased (decreased) in size by 10%. **i** Transmission simulation for varying defect density. For a defect density of 50% the bandgap is clearly distinguishable in the transmission plots.

**Supplementary Note 7: Phononic band gap vs. layer number**

As a potential approach to overcome challenges associated with the fabrication of uniform and residue free suspended graphene devices for phononic pattering, we suggest to use thin multilayers. For this we have to check, if the bandgap persist also for thicker devices. We thus perform band structure calculations for various thickness (maintaining a constant stress in the device) and extract the band gap (Supplementary Figure 7). We find a bandgap up to ~350 layers of graphene.

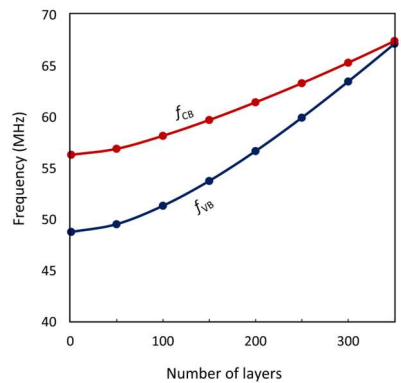

**Supplementary Figure 7 | Phononic band gap vs. layer number.** Valence band maximum ($f_{VB}$) and conduction band minimum ($f_{CB}$) vs. number of graphene layers extracted from band structure calculations.

**Supplementary Note 8: Localization-delocalization transition**

To further highlight the usefulness of the proposed system and to strengthen the analogy to a MIT, we perform an additional study. In it, we examine the localization of defect vibrational modes that can be compared to the localization of mid-gap defect states in semiconductors. By varying the size of the bandgap using our tension engineering approach, we observed the behaviour similar to the localization-delocalization transition in solids.

In this study, we place an artificial lattice defect within the phononic lattice. When the defect is within a phononic bandgap, it hosts spatially localized vibrational modes. We then simulate the pressure dependence of such a defective phononic lattice using two geometries. The first device is similar to the one shown in the manuscript (Supplementary Figure 8a), where we expect the bandgap closing under the application of pressure. The second one is a circular reference device (Supplementary Figure 8b, as studied in previous work). As explained in the main text, the bandgap does not close for this geometry. At zero pressure, the defect mode is localized in both geometries (Supplementary Figure 8). At the same time, we find starkly different behaviour of that mode in circular and rectangular devices under applied pressure (Supplementary Figure 8c,d). In the rectangular device, the mode starts to show displacement over the entire device (highlighted by grey arrows) and thus loses its localization as the bandgap closes. In the circular device, in contrast, the bandgap does not close, and the defect mode always stays within it. Correspondingly, the defect mode remains localized in the entire range of pressures (Supplementary Figure 8d). This further highlights the similarity between phononic and solid-state crystals, despite their very different quasiparticles. We believe that the behaviour we observe can be compared to the localization-delocalization transition.

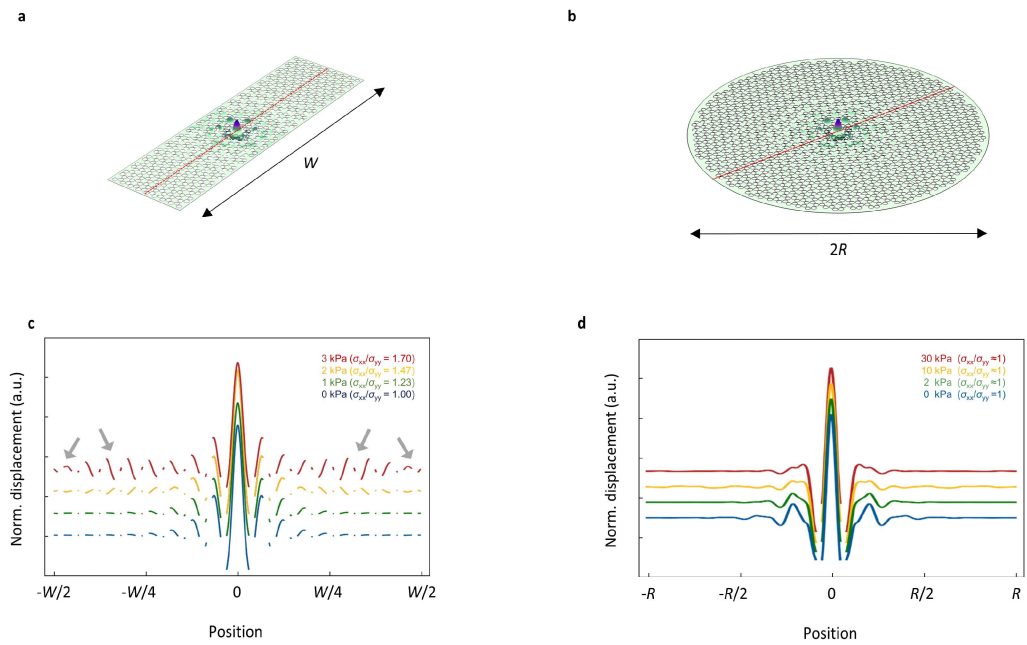

**Supplementary Figure 8 | Dephasing a localized defect state using uniaxial tension. a,b** Mode shape of a localized mode within the bandgap of a rectangular (a) and circular (b) phononic device **c**, Line cut of the normalized displacement extracted along the red line in **a** vs. applied pressure (plots are offset for better visibility). With increased pressure, the degree of uniaxiality is increased and the bandgap gradually closes. At the same time the mode that was localized near device centre at zero pressure becomes delocalized over the entire device. **d,** Same as **c**, but for the circular device shown in (b). Here the mode shape remains virtually unchanged under pressure, as the frequency of the defect mode scales together with the bandgap and maintains its localization.

# Supplementary References

1. Kirchhof, J. N. *et al.* Tunable Graphene Phononic Crystal. *Nano Lett.* **21**, 2174–2182 (2021).